\documentclass[aps,prd,showpacs,showkeys,amsmath,amssymb]{revtex4}
\usepackage{graphicx}
\begin{document}

\title{Possible Interpretations of $D_{sJ}^+(2632)$ If It {\sl Really} Exists}
\author{Yuan-Ben Dai}
\author{Chun Liu}
\affiliation{Institute of Theoretical
Physics,  Chinese Academy of Sciences, P.O. Box 2735, Beijing
100080,  China}
\author{Y.-R. Liu}
\author{Shi-Lin Zhu}
\email{zhusl@th.phy.pku.edu.cn} \affiliation{Department of
physics, Peking University, Beijing 100871, China}

\date{\today}

\begin{abstract}

We analyze various possible interpretations of the narrow state
$D_{sJ}^+(2632)$ observed by SELEX Collaboration recently, which
lies above threshold and has abnormal decay pattern. These
interpretations include: (1) several versions of tetraquarks; (2)
conventional $c\bar s$ meson such as the first radial excitation
of $D_s(2112)$ with abnormally large $SU(3)$ symmetry breaking;
(3) conventional $c\bar s$ meson with abnormally large $\eta_1$
coupling; (4) heavy hybrid meson. We discuss the physical
implications of each interpretation. For example, if the existence
of $D_{sJ}^+(2632)$ is confirmed as the first radial excitation of
$D_s(2112)$ by other experiments, it will be helpful to look for
(1) its SU(3) flavor partners $D_{J}^{0,+}(2530)$; (2) its B-meson
analogues $B_{J}^{0,+}(5840), B_{sJ}^+(5940)$; (3) S-wave two pion
decay modes.

\end{abstract}

\maketitle

\section{Difficulty with $c\bar s$ assignment
and SU(3) flavor symmetry}\label{sec1}

The experimental discovery of narrow low-lying charm-strange
mesons $D_{sJ}(2317), D_{sJ}(2457)$ is an important event in heavy
meson spectroscopy \cite{babar,cleo,belle1,belle2}. Many
theoretical interpretations were proposed
\cite{cheng,rev1,bardeen,dai,rev2}. However, the large
electromagnetic branching ratio of $D_{sJ}(2457)$ observed by
BELLE Collaboration favors the assignment of these two states as
conventional $c\bar s$ states \cite{bardeen,dai,rev2}. These two
states belong to the $(0^+, 1^+)$ doublet with $j_l={1\over 2}^+$
in the heavy quark effective field theory.

Recently SELEX Collaboration observed an exotic charm-strange
meson $D_{sJ}(2632)$.  Its decay width is very narrow, $\Gamma <
17$ MeV at $90\%$ C.L.  The decay channels are $D_s \eta$ and
$D^0K^+$ with the unusual relative branching ratio:
$\Gamma(D^0K^+)/\Gamma(D_s\eta)=0.16\pm 0.06$ \cite{selex}. This
observation has inspired a lot of theoretical papers
\cite{maiani,chen,nicolescu,chao,barnes,liu,nowak,beveren,gupta,zhang}.

In order to explain why this relative branching ratio is abnormal,
let's first assume (1) SU(3) flavor symmetry and (2) that the
quark content of $D_{sJ}^+(2632)$ is $c\bar{s}$. The charge
conjugate state $D_{sJ}^-(2632)$ belongs to $SU(3)_F$ triplet when
the heavy quark transforms as a $SU(3)_F$ singlet. We denote the
triplet as
\begin{equation}
 (T^i)=\bordermatrix{&\cr & \bar{c}u\cr& \bar{c}d\cr
&\bar{c}s}=\bordermatrix{&\cr & \bar{D}^0_J\cr& \bar{D}_J^-\cr &
\bar{D}_{sJ}(2632)}
\end{equation}
where $\bar{D}^0_J, \bar{D}_J^-$ are SU(3) flavor partners of
$\bar{D}_{sJ}(2632)$.

Then the decay of heavy meson triplet $T^\prime$ can be described
with the effective Lagrangian in the $SU(3)_F$ favor symmetry
limit
\begin{equation}\label{lag}
L_8=g_8 T^{\prime \dagger}_i M^i_j T^j+ \mbox{H.C.}
\end{equation}
where $M^i_j$ is the matrix of pseudoscalar meson octet
\begin{eqnarray}
(M^i_j) &=&\left(\begin{array}{ccc}
\frac{\pi^0}{\sqrt{2}}+\frac{\eta_8}{\sqrt{6}}&\pi^+&K^+\\
 \pi^-&-\frac{\pi^0}{\sqrt{2}}+\frac{\eta_8}{\sqrt{6}}&K^0\\
K^-&\bar{K}^0&-\frac{2\eta_8}{\sqrt{6}} \end{array}\right)\; ,
\end{eqnarray}
and $T^j$ is the final state heavy pseudoscalar meson triplet. We
suppressed the Lorentz indices for mesons.

The two-body decay width of a meson reads
\begin{equation}\label{ratio}
\Gamma=g^2\frac{k^{2L+1}}{m_0^{2L}},
\end{equation}
where $g$ is the dimensionless effective coupling constant, $L$ is
the angular momentum for decay. $m_0$ is the parent mass and $k$
is the decay momentum in the center of mass frame
\begin{equation}
k=\frac{1}{2m_0}
\{[m_0^2-(m_1+m_2)^2][m_0^2-(m_1-m_2)^2]\}^{\frac12},
\end{equation}
where $m_1$ and $m_2$ are the masses of final mesons. The ratio of
decay widths of these two channels for $D^+_{sJ}(2632)$ is
\begin{equation}
\frac{\Gamma (D^0 K^+)}{\Gamma (D_s^+\eta)}=(\frac{g_{D^0
K^+}}{g_{D_s^+\eta}})^2(\frac{k_{D^0 K^+}}{k_{D_s^+\eta}})^{2L+1}.
\end{equation}

Using $\frac{\lambda_{D^0 K^+}}{\lambda_{D_s\eta}}=\sqrt{3\over
2}$, $k_{D^0 K^+}=499$ MeV, $k_{D_s\eta}=325$ MeV, we get
\begin{equation}
\frac{\Gamma (D^0 K^+)}{\Gamma (D_s\eta)}=2.3*(1.54)^{2L}\ge 2.3.
\end{equation}
which is nearly 15 times larger than the experimental value
\cite{selex}!

Clearly either some important physics is missing or one of the
above two assumptions of SU(3) symmetry and $D_{sJ}(2632)$ as a
$c\bar s$ state is wrong. In the following sections we analyze
four different interpretations of this charming state and the
resulting phenomenology.

\section{Tetraquarks}\label{sec5}

One intriguing possibility is that $D_s(2632)$ is a tetraquark
\cite{maiani,chen,nicolescu,chao,liu,gupta}. Then, there should be
other tetraquark partners of $D_{sJ}(2632)$. Moreover, its narrow
width may require that quarks inside $D_{sJ}(2632)$ form tightly
bound clusters like diquarks \cite{chao}. We give a brief review
of different versions of tetraquark interpretations.

Tetraquarks with quark content $\bar c\bar q qq$ form four
multiplets: two triplets, one anti-sextet and one 15-plet
\begin{equation}
3\otimes3\otimes\bar{3}\otimes1=3_1\oplus3_2\oplus\bar{6}\oplus15
\; .
\end{equation}
The wave functions of these states can be found in Ref.
\cite{liu}. The identification of $D_{sJ}(2632)$ as the $J^P=0^+$
isoscalar member of the ${\bf 15}$ tetraquarks with the quark
content ${1\over 2\sqrt{2}}
(ds\bar{d}+sd\bar{d}+su\bar{u}+us\bar{u}-2ss\bar{s})\bar{c}$ leads
to the relative branching ratio \cite{liu}
\begin{equation}
{\Gamma(D_{sJ}^+(2632)\rightarrow {D}^0 K^+)\over
{\Gamma(D_{sJ}^+(2632)\rightarrow D_s^+\eta)}} =0.25\; .
\end{equation}
This decay pattern arises from the SU(3) Clebsch-Gordan
coefficients very naturally.

Another possibility is that $D^+_{sJ}(2632)$ is dominated by
$c\bar ss\bar s$ with $J^P=0^+$ \cite{chen,gupta}. The $s\bar s$
fluctuates into ${1\over \sqrt{3} \eta_1} -{2\over
\sqrt{6}\eta_8}$. Hence, its decay mode is mainly $D^+_s \eta$.
The final states $D^0K^+, D^+K^0$ are produced through the
annihilation of $s\bar s$ into $u\bar u+d\bar d$ which requires
$\eta_1$ component. Thus this process is OZI suppressed. In this
way the anomalous decay pattern is achieved.

In the diquark correlation configuration \cite{chao},
$D^+_{sJ}(2632)$ is suggested to be a $(cs)_{3^*}-(\bar s\bar
s)_3$ state where the subscript numbers are color representations.
With the assumption that $(cs)_{3^*}-(\bar s\bar s)_3$ has small
mixing with $(c\bar s)_1-(s\bar s)_1$, one can give a nice
interpretation for the narrow width. The mixing between $s\bar s$
and $u\bar u+d\bar d$ can lead to the unusual branching ratio.

All the above three tetraquark interpretations predict the same
production rates for $D^+K^0, D^0K^+$ final states. A serious
challenge is that SELEX Collaboration didn't find any signal in
the $D^+K^0$ channel \cite{selex}.

A very different tetraquark version $cd\bar d\bar s$ is proposed
in Refs. \cite{maiani,nicolescu}. Naively, one would expect
$D^0K^+$ decay channel is suppressed while $D^+ K^0$ and $D_s
\eta$ modes are both important. Interestingly, Ref. \cite{maiani}
invoked the isospin symmetry breaking to explain the relative
branching ratio. It was proposed that the mass eigenstate
$D^+_{sJ}(2632)$ is the mixture between the two flavor eigenstates
$a^+_{c\bar s}$ and $f^+_{c\bar s}$, where $a^+_{c\bar s}$ is the
$I=1$, $I_3=0$ state in $SU(3)_F$ 6 representation and $f^+_{c\bar
s}$ is the $I=0$ state in $\bar{3}_1$ with our notation. With some
special mixing scheme, the relative ratio is found to be
${\Gamma(D^0K^+)\over \Gamma(D^+_s\eta)}=0.16$. At the same time,
the authors of Ref. \cite{maiani} predicted $4<
{\Gamma(D^+K^0)\over \Gamma(D^+_s\eta)}<7.6$, $1.7< {\Gamma(D_s
\pi^0)\over \Gamma(D^+_s\eta)}<6.5$.

\section{Large $SU(3)_F$ Breaking and $D_{sJ}(2632)$ AS a
$c\bar s$ state}\label{sec2}

Some authors suggested that $D_{sJ}(2632)$ be the first radial
excitation of $D_s(2112)$ with $J^P=1^-$. The nodal structure of
the radial wave function of $D_s(2632)$ ensures the narrow width
while different decay momentum in two channels lead to anomalous
decay pattern \cite{chao,barnes}.

We want to emphasize that two new narrow states $D^0_J$ and
$D^+_J$ should also exist as flavor partners of $D_{sJ}(2632)$
within this scheme. The mass difference between $D_{sJ}(2632)$ and
$D^0_J$ should roughly be the strange quark mass $m_s=100$ MeV .
So their masses are around 2532 MeV. If $D_{sJ}(2632)$ is
interpreted as the first radial excitation of $D_s(2112)$, $D_J^0
(2532)$ should be the first radial excitation of $D^\ast (2010)$.
Similarly one expects the following B-meson analogues:
$B_{J}^{0,+}(5840), B_{sJ}^+(5940)$. Moreover, the branching ratio
of the S-wave two pion decay modes may be significant if
$D_{sJ}(2632)$ is the radial excitation.

We want to point out that the decay momentum of the two channels
$D^+_s \eta$ and $D^0K^+$ are the same in the exact $SU(3)_F$
symmetry. Then there is no anomalous decay pattern. In other
words, the origin of this kind of explanation of the anomalous
decay pattern can be traced back to the SU(3) falvor symmetry
breaking effects. In the following we shall use the effective
Lagrangian formalism to analyze this effect in a model-independent
way.

The symmetry breaking is caused by the quark mass matrix
$m=diag(\hat{m}, \hat{m}, m_s)$. We assume the isospin symmetry.
The symmetry breaking Lagrangian reads
\begin{eqnarray}
L_m&=&\alpha T^{\prime \dagger}_i m^i_j (S_3)^j +\beta T^{\prime
\dagger}_i m^l_j (S_{\bar{6}})_{lk}\epsilon^{ijk} +\gamma
T^{\prime \dagger}_i m^k_j (S_{15})^{ij}_k
\end{eqnarray}
where
\begin{eqnarray}
(S_3)^i&=&M^i_j T^j,\nonumber\\
(S_{\bar{6}})_{ij}&=&M^a_j T^b \epsilon_{iab} +M^a_i T^b
\epsilon_{jab}\nonumber
\end{eqnarray}
and
\begin{eqnarray}
(S_{15})^{ij}_k=M^i_k T^j + M^j_k T^i -\frac{1}{4}(\delta^i_k
M^j_a T^a + \delta^j_k M^i_a T^a).\nonumber
\end{eqnarray}

For $D_{sJ}^-(2632)$, we have
\begin{eqnarray}
L_{D_{sJ}}&=&\{(g_8+\alpha m_s)-\frac12\gamma
(m_s-\hat{m})\}(D^+_{sJ}K^-\bar{D}^0+D^+_{sJ}\bar{K}^0D^-)\nonumber\\
&&-\frac{2}{\sqrt6}\{(g_8+\alpha m_s)+\frac32\gamma(m_s-
\hat{m})\}D^+_{sJ}\eta_8 D^-_s.
\end{eqnarray}
\begin{eqnarray}
\frac{g_{\bar
D^0K^-}}{g_{D^-_s\eta_8}}=\frac{1}{\sqrt6}[1-\frac{4(g_8+\alpha
m_s)}{(g_8+\alpha m_s)+\frac32\gamma(m_s-\hat m)}].
\end{eqnarray}
Naively one would expect $|\frac{\gamma(m_s-\hat m)}{g_8+\alpha
m_s}|\ll 1$. If we assume the physical $\eta$ meson is
approximately $\eta_8$, we can extract the value
$\frac{\gamma(m_s-\hat m)}{g_8+\alpha m_s}$ from the relative
branching ratio $\frac{\Gamma (D^0 K^+)}{\Gamma (D_s^+\eta)}$.

Numerically there are two possibilities:
\begin{equation}\label{posi}
\frac{\gamma(m_s-\hat m)}{g_8+\alpha m_s}=4.8 \;.
\end{equation}
Or
\begin{equation}\label{nega}
\frac{\gamma(m_s-\hat m)}{g_8+\alpha m_s}=1.1 \;.
\end{equation}
Although the above numbers seem quite unnatural, the anomalously
large $SU(3)_F$ breaking can explain the special decay pattern of
$D_{sJ}(2632)$ in principle.

For $\bar{D}^0_J$ and $D^-_J$,
\begin{eqnarray}
L_{D_J}&=&\{(g_8+\alpha \hat{m})-(\beta+\frac14\gamma)
(m_s-\hat{m})\}\{(\frac{1}{\sqrt2}D^0_{J}\pi^0\bar{D}^0+D^0_{J}\pi^+D^-)\nonumber\\
&&+(D^+_{J}\pi^-\bar{D}^0-\frac{1}{\sqrt2}D^+_{J}\pi^0 D^-)\}
+\{(g_8+\alpha \hat{m})\nonumber\\
&&+(\beta+\frac34\gamma)
(m_s-\hat{m})\}(D^0_JK^+D^-_s+D^+_JK^0D^-_s)\nonumber\\
&&+\frac{1}{\sqrt6}\{(g_8+\alpha
\hat{m})+3(\beta-\frac34\gamma)(m_s- \hat{m})\}(D^0_{J}\eta_8
\bar{D}^0+D^+_{J}\eta_8 D^-)\; .
\end{eqnarray}

In general, $D^{0,+}_J$ may not have the same decay pattern as
$D_{sJ}(2632)$. For example, in the extreme case $|\alpha|\ll
|\gamma|, |\beta|\ll |\gamma|$, we have the following relative
branching ratios
\begin{equation}
\Gamma ( D^+_J\to D^0\Pi^+): \Gamma ( D^+_J\to D^+\Pi^0): \Gamma (
D^+_J\to D_s^+ {\bar K}^0):\Gamma ( D^+_J\to D^+\eta)=1:0.5: 36.6:
74.8
\end{equation}
for the ratio in Eq. (\ref{posi}) and
\begin{equation}
\Gamma ( D^+_J\to D^0\Pi^+): \Gamma ( D^+_J\to D^+\Pi^0): \Gamma (
D^+_J\to D_s^+ {\bar K}^0):\Gamma ( D^+_J\to D^+\eta)=1:0.5: 0.4:
0.1
\end{equation}
for the ratio in Eq. (\ref{nega}).

\section{Large Singlet Coupling and $D_{sJ}(2632)$ AS a
$c\bar s$ state}\label{sec3}

As in the last section we assume $D_{sJ}(2632)$ is a $c\bar s$
state with $J^P=1^-$. Instead of invoking very large SU(3)
breaking effects to account for its decay pattern, we introduce
the interaction between SU(3) singlet $\eta_1$ and $D_{sJ}(2632)$
and consider the mixing between $\eta_8$ and $\eta_1$. Now the
effective Lagrangian becomes
\begin{equation}
L_{eff}=g_8 T^{\prime \dagger}_i M^i_j T^j + g_1 \eta_1 T^{\prime
\dagger}_i T^i.
\end{equation}

The physical states $\eta$ and $\eta^\prime$ are admixtures of
$\eta_8$ and $\eta_1$,
\begin{eqnarray}
\eta&=&\eta_8\cos\theta-\eta_1\sin\theta \nonumber\\
\eta^\prime&=&\eta_8\sin\theta+\eta_1\cos\theta
\end{eqnarray}
with the mixing angle $\theta= -20^\circ$ \cite{pdg}. Now we have
\begin{eqnarray}
L_{D_{sJ}}&=&g_8\{D_{sJ}^{+} K^- \bar{D}^0 + D_{sJ}^{+} \bar{K}^0
D^- - (\frac{2}{\sqrt6}\cos\theta + \frac{g_1}{g_8} \sin \theta)
D_{sJ}^{+} \eta D_s^- \nonumber\\
&&+ (-\frac{2}{\sqrt6}\sin\theta + \frac{g_1}{g_8} \cos \theta)
D_{sJ}^{+} \eta^\prime D_s^- \}.
\end{eqnarray}
The Lagrangian involving $\bar{D}^0_J$ and $D^-_J$ reads
\begin{eqnarray}
L_{D_J}&=&g_8\{\frac{1}{\sqrt2}D^0_J\pi^0\bar{D}^0 + D^0_J \pi^+
D^- + D^0_J K^+ D^-_s\nonumber\\
&&- \frac{1}{\sqrt2}D^+_J\pi^0 D^- +
D^+_J \pi^- \bar{D}^0 + D^+_J K^0 D^-_s \nonumber\\
&&+(\frac{1}{\sqrt6}\cos\theta - \frac{g_1}{g_8} \sin \theta)(
D^0_J \eta \bar{D}^0+ D^+_J \eta D^-)\nonumber\\
&&+ (\frac{1}{\sqrt6}\sin\theta +\frac{g_1}{g_8} \cos \theta)
(D^0_J \eta^\prime \bar{D}^0+D^+_J \eta^\prime D^-)\} .
\end{eqnarray}

From the relative decay ratio we can extract either
\begin{equation}\label{posit}
\frac{g_1}{g_8}=16.2
\end{equation}
or
\begin{equation}\label{negat}
\frac{g_1}{g_8}=-11.7 \;.
\end{equation}
With these values, we have following relative branching ratios
\begin{equation}
\Gamma ( D^+_J\to D^0\Pi^+): \Gamma ( D^+_J\to D^+\Pi^0): \Gamma (
D^+_J\to D_s^+ {\bar K}^0):\Gamma ( D^+_J\to D^+\eta)=1:0.5: 0.07:
6.6
\end{equation}
for the ratio in Eq. (\ref{posit}) and
\begin{equation}
\Gamma ( D^+_J\to D^0\Pi^+): \Gamma ( D^+_J\to D^+\Pi^0): \Gamma (
D^+_J\to D_s^+ {\bar K}^0):\Gamma ( D^+_J\to D^+\eta)=1:0.5: 0.07:
2.4
\end{equation}
for the ratio in Eq. (\ref{negat}).

It is interesting to note that $\eta_1$ mixes strongly with
$G\tilde{G}$ because of axial anomaly. The anomalously large
coupling between $\eta_1$ and $D_{sJ}^+(2632)$ indicates that
$D_{sJ}^+(2632)$ may be a heavy hybrid meson containing explicit
glue \cite{zhu}. That's the subject of the next section.

\section{Large $N_c$ limit and $D_{sJ}^+(2632)$ as a heavy hybrid meson}\label{sec4}

\begin{figure}
\scalebox{0.9}{\includegraphics{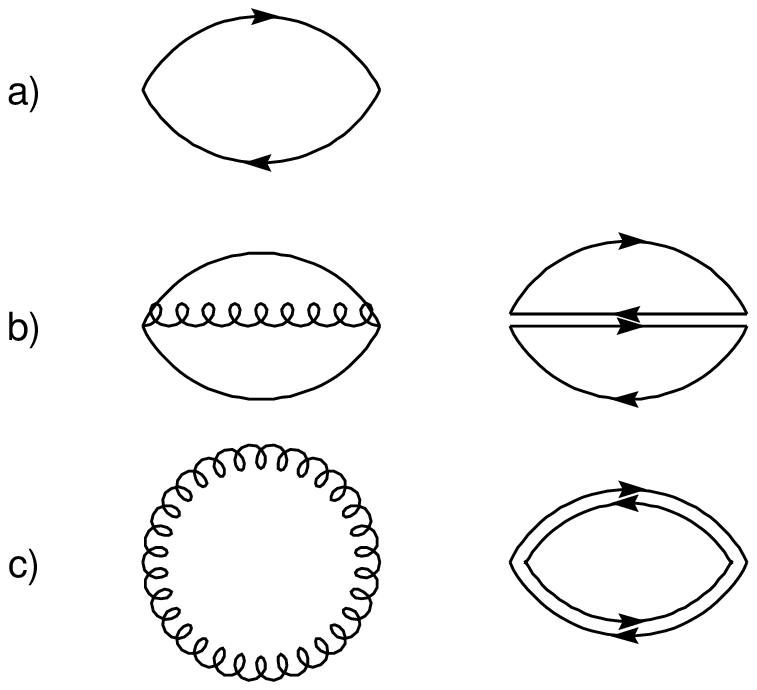}} \vspace{0.5cm}
\caption{Feynman diagrams and the corresponding double-line
representation for the decay modes of $J^P=1^-$ heavy hybrid
meson. } \label{fig1}
\end{figure}

Large $N_c$ formalism is a very useful and unique tool which is
applicable in the whole energy regime from zero to infinity. In
this section, we explore the possibility of $D_s(2632)$ being a
hybrid state containing explicit glue using large $N_c$ expansion.
Recall that the normalized interpolating currents for a pure
glueball and conventional meson A in the large $N_c$ limit read
\begin{equation}
O_A ={1\over \sqrt{N_c}} {\bar q}^i \Gamma_A q^i
\end{equation}
\begin{equation}
O_G =g_s^2 G^2 \; .
\end{equation}
The factor ${1\over \sqrt{N_c}}$ is introduced to ensure the
creation amplitude by $O_A$ from the vacuum is $\sim {\cal
O}(N_c^0)$ in the large $N_c$ limit. I.e., the following matrix
elements are order unity when $N_c\to \infty$:
\begin{equation}
\langle |0|O_A(0)|A\rangle = F_A \sim {\cal O}(N_c^0)
\end{equation}
\begin{equation}
\langle |0|O_G(0)|\mbox{Glueball}\rangle = F_G \sim {\cal
O}(N_c^0) \; .
\end{equation}
We have depicted $q\bar q$ mesons, $\bar q Gq$ hybrid states and
glueballs in Fig. \ref{fig1}.

\begin{figure}
\scalebox{1.0}{\includegraphics{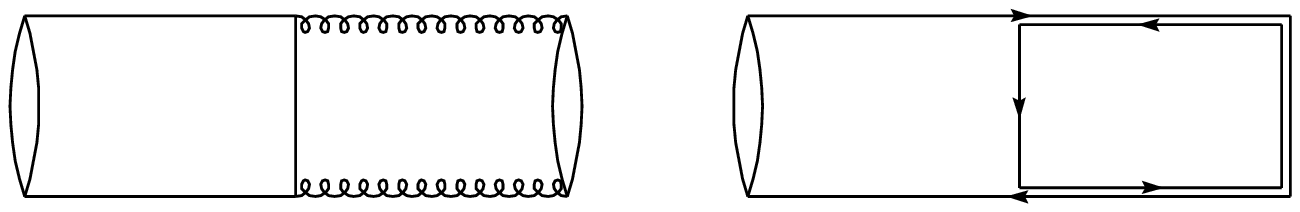}} \vspace{0.5cm}
\caption{Feynman diagrams and the corresponding double-line
representation for the decay modes of $J^P=1^-$ heavy hybrid
meson. } \label{fig2}
\end{figure}

In order to illustrate the formalism, we consider the $N_c$ order
of the mixing amplitude between a glueball and $q\bar q$ meson,
which can be extracted through the correlation function:
\begin{equation}
\langle 0| {\hat T} O_A(x) O_G(y) |0\rangle \; .
\end{equation}
The Feynman diagram is presented in Fig. \ref{fig2}. Also shown is
the double-line representation in the large $N_c$ approach. There
are two independent closed color loops which contribute $N_c^2$
while two vertices contribute $g_s^2$. Its $N_c$ order reads
\begin{equation}
\langle 0| {\hat T} O_A(x) O_G(y) |0\rangle \sim {1\over
{\sqrt{N_c}}} g_s^4 N_c^2= N_c^{-{1\over 2}}; .
\end{equation}
In other words, glueballs decouple from the conventional mesons
when $N_c\to \infty$ \cite{witten}.

We consider two kinds of heavy hybrid mesons: (1)
\begin{equation}
O_{H_1}={1\over \sqrt{N_c}}\bar s \gamma^\nu g_s G_{\mu\nu} c
\end{equation}
with $J^P=1^-$ and $c\bar s$ in the color octet state or (2)
\begin{equation}
O_{H_2}={1\over \sqrt{N_c}}\bar s i\gamma_5 c
g_s^2G_{\mu\nu}{\tilde G}^{\mu\nu}
\end{equation}
with $J^P=0^+$ and $c\bar s$ in the color singlet state.
\begin{equation}
<0|O_{H_{1\mu}}(0)|\mbox{Vector Hybrid State}:  H_1> =
F_{H_1}\epsilon_\mu \sim {\cal O}(N_c^0)
\end{equation}
\begin{equation}
<0|O_{H_2} (0)|\mbox{Scalar Hybrid State}: H_2> = F_{H_2} \sim
{\cal O}(N_c^0) \;.
\end{equation}

\begin{figure}
\scalebox{0.8}{\includegraphics{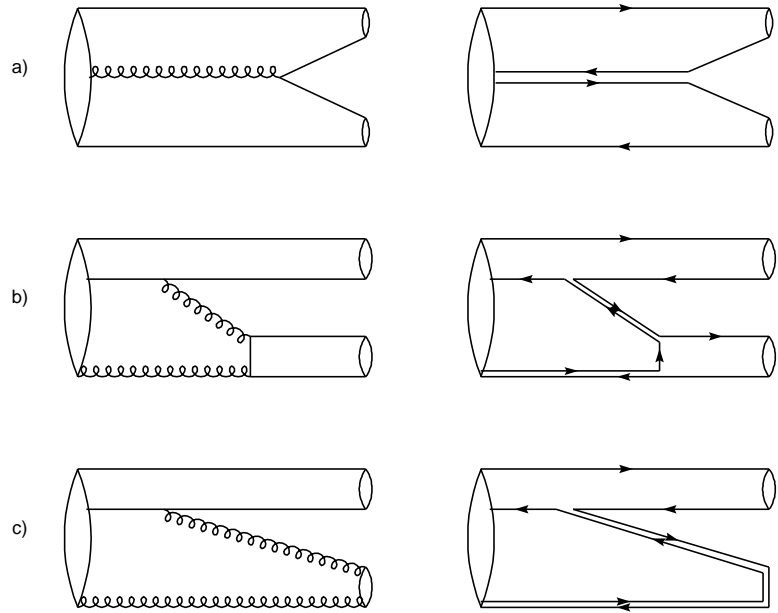}} \vspace{0.5cm}
\caption{Feynman diagrams and the corresponding double-line
representation for the decay modes of $J^P=1^-$ heavy hybrid
meson. } \label{fig3}
\end{figure}

For the $J^P=1^-$ hybrid meson, its decay channels are shown in
Fig. \ref{fig3} using the Feynman diagram and double-line
representation. The dominant decay mode in the large $N_c$ limit
is for the gluon to split into a color-octet $q\bar q$ pair in
diagram (a) in Fig. \ref{fig3}. Then the $q\bar q$ recombine with
$c\bar s$ pair to form $D^0K^+, D^+ K^0, D_s\eta$. The $N_c$ order
of this decay amplitude is $\sim \left({1\over {\sqrt
N_c}}\right)^3 g_s^2 N_c^2 =N_c^{-{1\over 2}}$. When it decays
into a glueball and $q\bar q$ meson, the decay amplitude is ${\cal
O}(N_c^{-1})$, i.e., diagram (c). When it decays into two $q\bar
q$ mesons through color-singlet two gluon intermediate states in
diagram (b), the decay amplitude is ${\cal O}(N_c^{-{3\over 2}})$.
In other words, the assignment of $D_{sJ}(2632)$ as a $J^P=1^-$
hybrid meson will not lead to the abnormal decay pattern observed
by SELEX Collaboration.

\begin{figure}
\scalebox{0.8}{\includegraphics{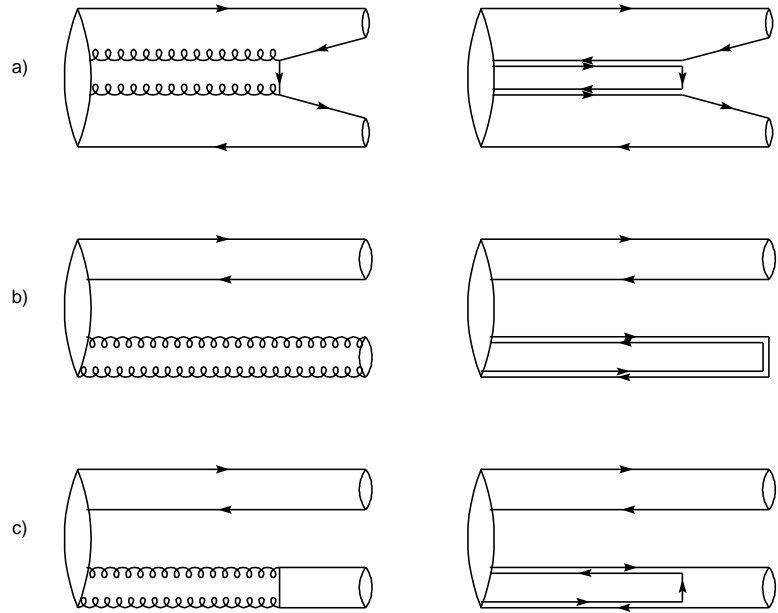}} \vspace{0.5cm}
\caption{Feynman diagrams and the corresponding double-line
representation for the decay modes of $J^P=0^+$ heavy hybrid
meson. } \label{fig4}
\end{figure}

Now we turn to the second case. With this configuration, it is
possible to understand the narrowness of $D_s(2632)$. When it
decays into a glueball and $D_s$ in diagram (b) in Fig.
\ref{fig4}, the amplitude is ${\cal O}(N_c^0)$ in the large $N_c$
limit. Hence, this would be the most possible decay channel if
kinematics allows. However, it cannot occur because the glueball
is too heavy.

The decay mode $D_s\eta_1$ in diagram (c) in Fig. \ref{fig4} is
${\cal O}(N_c^{-{1\over 2}})$ and subleading. However, it is also
forbidden by kinematics. Therefore $D_s(2632)$ is narrow. We note
that $\eta$ meson is a mixture of $\eta_8$ and $\eta_1$. The
mixing amplitude is ${\cal O}(N_c^0)$ arising from SU(3) symmetry
breaking effect. Hence the kinematically allowed decay mode $D_s
\eta$ becomes the dominant mode although it is ${\cal
O}(N_c^{-{1\over 2}})$.

$DK$ channel can happen only when the color-singlet two gluons
annihilate into a pair of light quarks, namely after the hybrid
becomes a four-quark state. This hybrid to four-quark transition
is $1/N_c$ suppressed in amplitude.  The four-quark state can
decay to both $D_s\eta$ and $DK$, and the later channel is further
suppressed compared to the former due to the color mismatch factor
which is $1/ {N_c}$ in amplitude. The $N_c$ order can be read from
diagram (a) in Fig. \ref{fig4}
\begin{equation}
\sim ({1\over \sqrt{N_c}})^3 g_s^4 N_c^2 =N_c^{-{3\over 2}} \;.
\end{equation}
Naively there are three closed color loops, hence one would get
the factor $N_c^3$. However, the charm and strange quark are in
the color singlet state. Hence their color indices are the same.
In other words, there are only two independent color loops.
Finally we obtain
\begin{equation}
{\Gamma(D^0K^+)\over \Gamma(D_s\eta)}\sim \frac{{\cal
O}(1/N_c^3)}{|{\cal O}(N_c^{-{1\over 2}}) +{\cal O}(N_c^{-{3\over
2}})|^2}={\cal O}(N_c^{-2})\sim {1\over 9} \times 1.5 \approx 0.16
\end{equation}
where the numerical value arises from using $N_c=3$. The
assignment of $D_{sJ}(2632)$ as a $c\bar s g_s^2 G \tilde G$ state
explains the decay pattern quite naturally. A dynamical
calculation of the mass of such a heavy hybrid meson will be very
desirable, which is beyond the present note.

\section{Discussions}\label{sec6}

In this note, we considered various possible interpretations of
$D^+_{sJ}(2632)$ and discussed the resulting phenomenology. Future
experimental confirmation of this state and discovery of its
partners will help unveil the mysterious underlying dynamics of
this narrow charm-strange meson.

S.L.Z. thanks Professor K.-T. Chao for helpful discussions. This
project was supported in part by the National Natural Science
Foundation of China, BEPC Opening Project, Ministry of Education
of China, FANEDD and SRF for ROCS, SEM.



\end{document}